\documentclass[sigconf]{acmart}

\usepackage{multirow}
\usepackage{enumitem}
\usepackage{booktabs}
\usepackage{graphicx}
\usepackage{natbib}
\usepackage{subfigure}
\newcommand\zc[1]{{\color{red} \textbf{ZC: }\color{blue}\textbf{#1}}}
\newcommand\gh[2]{{\color{red} \textbf{GH: }\color{blue}\textbf{#1}}}
\AtBeginDocument{%
  \providecommand\BibTeX{{%
    \normalfont B\kern-0.5em{\scshape i\kern-0.25em b}\kern-0.8em\TeX}}}

\begin{document}
\settopmatter{printacmref=false} 
\renewcommand\footnotetextcopyrightpermission[1]{} 
\author{Hao Guo}
\email{haoguo@tju.edu.cn}
\affiliation{%
  \institution{Tianjin Key Laboratory of Advanced Networking (TANK),  College of Intelligence and Computing}
\institution{Tianjin University}
  \city{Tianjin}
  \state{China}
}

\author{Zhenchang Xing}
\email{zhenchang.xing@anu.edu.au}
\affiliation{%
  \institution{Research School of Computer Science}
\institution{Australian National University}
  \state{Australia}
}

\author{Xiaohong Li}

\email{xiaohongli@tju.edu.cn}
\affiliation{%
  \institution{Tianjin Key Laboratory of Advanced Networking (TANK),  College of Intelligence and Computing}
\institution{Tianjin University}
  \city{Tianjin}
  \state{China}
}

\title{Predicting Missing Information of Key Aspects in Vulnerability Reports}
\begin{abstract}
	
	Software vulnerabilities have been continually disclosed and documented.
	An important practice in documenting vulnerabilities is to describe the key vulnerability aspects, such as vulnerability type, root cause, affected product, impact, attacker type and attack vector, for the effective search and management of fast-growing vulnerabilities.
	We investigate 120,103 vulnerability reports in the Common Vulnerabilities and Exposures (CVE) over the past 20 years.
	We find that 56\%, 85\%, 38\% and 28\% of CVEs miss vulnerability type, root causes, attack vector and attacker type respectively.
	To help to complete the missing information of these vulnerability aspects, we propose a neural-network based approach for predicting the missing information of a key aspect of a vulnerability based on the known aspects of the vulnerability.
	We explore the design space of the neural network models and empirically identify the most effective model design.
	Using a large-scale vulnerability datas\-et from CVE, we show that we can effectively train a neural-network based classifier with less than 20\% of historical CVEs.
	Our model achieves the prediction accuracy 94\%, 79\%, 89\%and 70\% for vulnerability type, root cause, attacker type and attack vector, respectively.
	Our ablation study reveals the prominent correlations among vulnerability aspects and further confirms the practicality of our approach.
	
\end{abstract}

\maketitle

\section{Introduction}
Software vulnerabilities are flaws or weaknesses present in software products, which can be exploited to damage system or information confidentiality, integrity and availability~\cite{cvss}.
Significant efforts have been made to document and manage publicly known software vulnerabilities.
At the core of these efforts is the Common Vulnerabilities and Exposures (CVE)~\cite{cve}.
CVE is a list of entries - each reporting a publicly known vulnerability with an unique identification number, a
description, and at least one public reference.
CVE entries are used or enriched in numerous security products and services, such as U.S. National Vulnerability Database\cite{nvd}, Symantec SecurityFocus~\cite{sf}, CVE Details~\cite{CVEdetail}.

\begin{figure}
	\centering

	\includegraphics[scale=0.470]{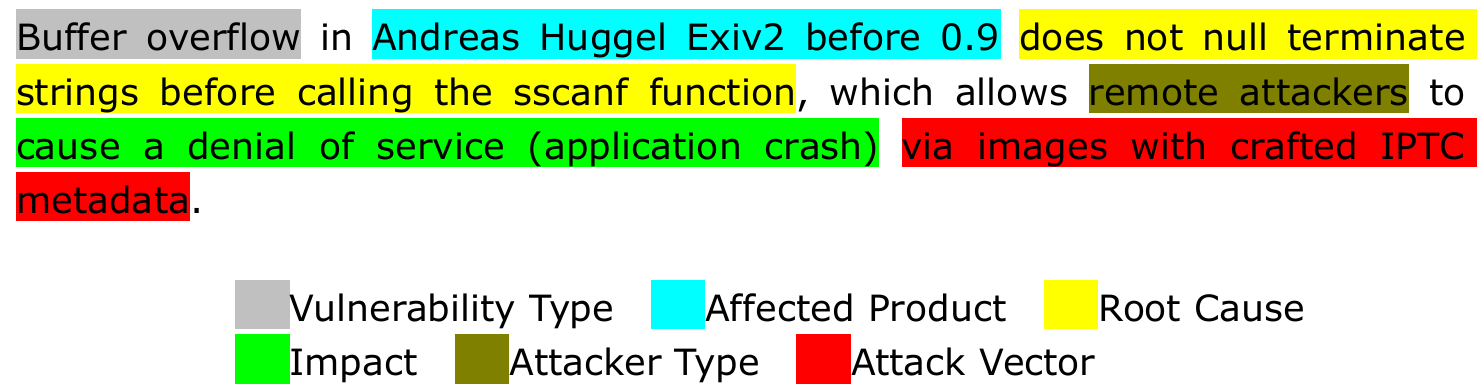}
\setlength{\abovecaptionskip}{0cm}
	\setlength{\belowcaptionskip}{0cm}

	\caption{An example of CVE description}
	\label{fig:cvedesccomplete}	
\end{figure}

\begin{figure}
\vspace{-4mm}

	\centering
	\subfigtopskip=0pt 
	\subfigbottomskip=0pt 
	\subfigcapskip=-5pt
  \subfigure[CVE-2019-5934: missing Root Cause]{\includegraphics[scale=0.460]{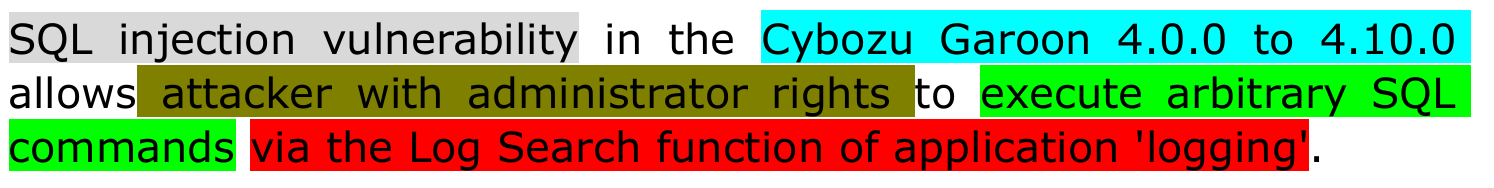}}

  \subfigure[CVE-2019-13561: missing Root Cause and Vulnerability Type]{\includegraphics[scale=0.460]{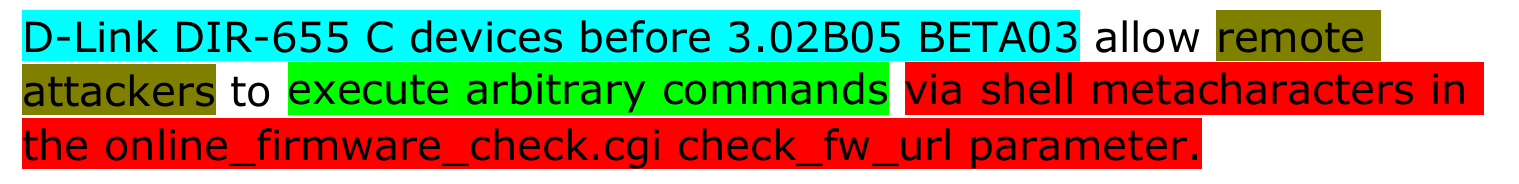}}

  \subfigure[CVE-2019-15117: missing Vulnerability Type]{\includegraphics[scale=0.460]{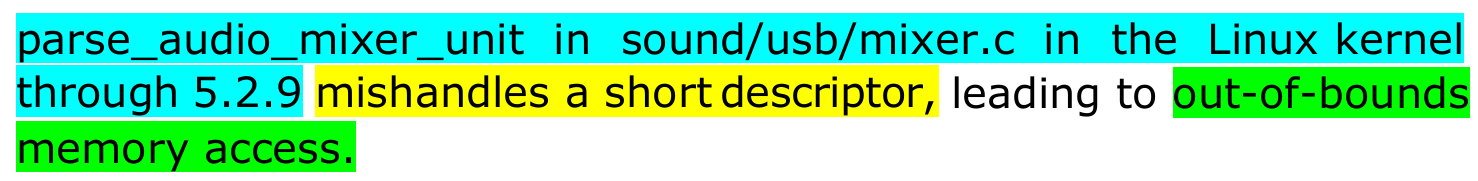}}

 \subfigure[CVE-2019-14262: only have Affected Product and Impact]{\includegraphics[scale=0.460]{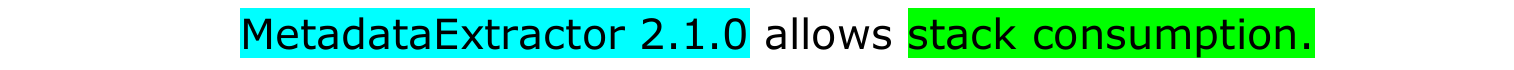}}

	\vspace{-3mm}
	\caption{Examples of CVEs that miss information}
	\vspace{-4mm}
	\label{fig:cvedescmissing}	
\end{figure}

Figure~\ref{fig:cvedesccomplete} shows the CVE description of the vulnerability entry CVE-2005-4676.
As highlighted in this example, a high-quality CVE description should contain six key aspects of details of the vulnerability~\cite{poj}, including \textit{vulnerability type} (e.g., buffer overflow), \textit{affected product} (often also include relevant vendor/version/compo\-nent information) (e.g., Andreas Huggel Exiv2 before 0.9), \textit{root cause} (e.g., does not null terminate strings before calling sscanf ), \textit{attacker type} (e.g., remote attacker), \textit{impact} (e.g., cause a denial of service (application crash)), and \textit{attack vector} (e.g., via images with crafted IPTC metadata).
Describing these key aspects is crucial for CVE management, mitigation and prevention, for example, detect and remove duplicate CVE entries~\cite{chen2018categorizing}, identify common weaknesses and attack patterns~\cite{cwe,capec}, timely analysis of high-risk CVEs~\cite{sf,ibm,exploit}.

We collect 120,103 CVEs from January 1999 to October 2019, and inspect the presence or absence of the above-mentioned six key aspects in the description of these 120,103 CVEs.
We find that almost all (over 99\%) of CVEs describe affected product.
In fact, when submitting a new CVE request, the reporter must specify the affected product.
However, all other five key aspects can be left unspecified.
Figure~\ref{fig:cvedescmissing} presents some examples.
Among these five key aspects, 94\% of CVEs describe the impact.
Impact describes what the attacker gains by exploiting this vulnerability, which is relatively easy to observe.
Compared with the impact, much more CVEs miss the other four more technical aspects - vulnerability type, root cause, attacker type and attack vector.
According to the guideline of CVE key details~\cite{poj}, a CVE should describe either vulnerability type or root cause, but not necessarily both.
We found that 58\% of CVEs describe either vulnerability type or root cause, and a small 3\% of CVEs describe both.
But the rest 42\% describe neither of them.
Furthermore, 28\% of CVEs miss attacker type, and 38\% of CVEs miss attack vector.

Some third-party communities or security product vendors may augment the missing details.
For example, CVE Details~\cite{CVEdetail} community attempt to identify vulnerability type of each CVE, and Security Focus~\cite{sf} tries to assign a root-cause class to each CVE.
However, there are often inconsistencies across different data sources~\cite{dong2019towards}.
We observe that as the coherent aspects of a vulnerability, different aspects of vulnerability manifest semantic correlations.
In this work, we would like to explore these semantic correlations to predict the missing aspects of a vulnerability based on its known aspects, without the dependence on third-party input.
Rule-based approach is infeasible in this task for two main reasons.
First, vulnerability aspects are described in natural language, and thus have large variations in terms of words and expressions used (see Table~\ref{tab:aspdes} for examples).
Second, the correlations among different vulnerability aspects and their combinations are intricate, and thus cannot be easily expressed as a set of explicit rules (see the experiment results of aspect fusion and ablation in Section~\ref{sec:aspectfusion} and Section~\ref{sec:aspectablation}).

In this work, we solve the prediction task using a neural-network based text classifier.
We carefully explore the design space of the neural-network based classifier, including two model architectures (early aspect fusion versus late aspect fusion), three types of input text (original CVE description, the concatenation of CVE aspects in original appearance order, and the concatenation of randomly shifted CVE aspects), general versus domain specific word embeddings, and two backbone networks (Convolutional Neural Network (CNN) versus Long-Short Term Memory (LSTM) with different network configurations (different network layers and with or without attention mechanism) (see Section~\ref{sec:approach}).
The neural-netwo\-rk based classifier learns the intricate correlations among different vulnerability aspects directly from the natural language description of existing vulnerabilities, and thus removes the need of manual feature engineering.
Once learned, these intricate correlations allow the classifier to predict the missing aspect of a vulnerability based on the known aspects of this vulnerability.

As over 94\% of CVEs do not miss affected product and impact, we focus the prediction on the other four aspects (vulnerability type, root cause, attacker type, and attack vector) which suffer from more serious information missing.
The prediction is a multi-class classification task.
Table~\ref{tab:freq} lists the class labels of the four aspects collected from 51,803 existing CVEs.
Section~\ref{sec:aspectextraction} describes how we collect these aspect labels.
We first randomly sample 20\% per year from 1999 to 2019. 
We extract CVE aspects from these sampled CVEs.
Based on the extracted CVE aspects, we develop regular expression patterns to extract CVE aspects from the rest of CVEs.
Considering the large number of the extracted CVE aspects, we adopt a sampling method~\cite{samp} to estimate the accuracy of the extracted information.
The extracted CVE aspects are of high quality ($>$97\% accuracy at the 95\% confidence level and 5\% error margin).

\begin{table}
	\vspace{-0.1cm}
	\setlength{\abovecaptionskip}{0.25cm}%
	\setlength{\belowcaptionskip}{0cm}%
	\caption{Variations of aspect descriptions}
	\vspace{-3mm}
	\label{tab:aspdes}
	\scriptsize
	\setlength{\tabcolsep}{0.9mm}{
		\begin{tabular}{clccc}
			\toprule
			Aspect&CVE-ID&Description\\
			\midrule
			\multirow{3}{*}{Vulnerability Type}&CVE-2017-11507&cross-site scripting (XSS) vulnerability \\
			&CVE-2018-16481&XSS vulnerability \\
			&CVE-2018-10937&cross site scripting flaw \\
			\midrule
			\multirow{3}{*}{Root cause} &CVE-2008-1419&does not properly handle ...\\
			&CVE-2010-0027&does not properly process ...\\
			&CVE-2015-1992&improperly processes ...\\
			\midrule
			\multirow{3}{*}{Affected product}&CVE-2006-3500&The dynamic linker (dyld) in Apple Mac OS X 10.4.7\\
			&CVE-1999-0786&The dynamic linker in Solaris \\
			&CVE-2013-0977&dyld in Apple iOS before 6.1.3 and Apple TV before 5.2.1 \\
			\midrule
			\multirow{3}{*}{Impact}&CVE-2011-4129&obtain sensitive information \\
			&CVE-2005-2436&obtain sensitive data\\
			&CVE-2002-0257&obtain information from other users\\
			\midrule
			\multirow{3}{*}{Attacker type} &CVE-2018-1000634&user with privilege\\
			&CVE-2018-1000084&low privilege user \\
			&CVE-2016-9603&A privileged user/process\\
			\midrule
			\multirow{3}{*}{Attack vector}&CVE-2018-12581&use a crafted database name\\
			&CVE-2019-11768& a specially crafted database name can be used\\
			&CVE-2012-1190&via a crafted database name\\
			
			\bottomrule
	\end{tabular}}
	\vspace{-5mm}
\end{table}

We build a dataset of 51,803 CVEs which contain at least four of the six CVE aspects. 
We use one aspect as the ``missing'' aspect to be predicted and the rest aspects as known aspects that the prediction is based on.
We conduct extensive experiments to study the impact of different design choices of the neural-network classifier on the prediction performance.
Our results show that the optimal network design should use early fusion model architecture and 1-layer CNN, and takes as input the concatenation of CVE aspects in the original appearance order and represents CVE aspect descriptions in CVE-specific word embeddings.
This optimal network design achieves the prediction accuracy 94\%, 79\%, 89\% and 70\% for vulnerability type, root cause, attacker type and attack vector, respectively.
We also conduct ablation experiments to determine the most and least prominent aspect or aspect combinations for predicting a particular missing aspect.
Our results show that impact aspect has the greatest impact on vulnerability type, while affected product and vulnerability type have the greatest impact on root cause, attacker type and attack vector.
At the same time, root cause and attacker type have least impact on the prediction of other aspects.

This paper makes the following contributions:
\begin{itemize}
	\item Our work is the first to investigate the aspect missing issue of the CVE entries. We analyze 24,042 CVEs over the 20 years to develop rules for extracting six CVE aspects from the CVE descriptions. We further analyze the severity and characteristics of different CVE aspects.
	\item We design a machine-learning approach for predicting the missing information of key aspects of CVEs based on the known aspects in the CVE descriptions. Our machine-learning model design systematically considers variations in input formats, word embeddings, model architectures and neural network designs.
	\item We conduct large-scale experiments to compare the effectiveness of different model design variants, investigate the prediction performance and the minimum effective amount of training data, and  identify the prominent correlations am\-ong different aspects for reliable prediction.
	
\end{itemize}

\section{Extraction of CVE Key Aspects}
\label{sec:aspectextraction}

In this section, we introduce the six key aspects for describing CVEs, discuss how we extract these key aspects from CVE descriptions, evaluate the quality of the extracted information, and report the statistics of different aspects absent in CVE descriptions.


\subsection{Preliminaries of CVE Aspects}
\label{sec:preliminary}

CVE suggests two description templates:
1) [Vulnerability Type] in [Component] in [Vendor][Product][Version] allows [Attacker Type] to [Impact] via [Attack Vector];
2) [Component] in [Vendor][Product][Version][Root Cause], which allows [Attacker Type] to [Impact] via [Attack Vector].
These two templates identify six key aspects for describing CVEs.
As the examples in Figure~\ref{fig:cvedescmissing} show, not all CVEs describe all six aspects.
When submitting a new CVE request, in addition to providing the natural language description of these six aspects, the submission form provides pre-defined options for vulnerability type, attacker type and impact. 
But the reporters may select Other if the pre-defined options are not appropriate for a CVE or leave them unselected.

\textbf{Vulnerability type} identifies an abstract software weakness that a specific CVE corresponds to. 
The abstract software weakness is usually identified as an entry in Common Weakness Enumeration (CWE)~\cite{cwe}.
When submitting a new CVE request, the CVE reporter must specify the vulnerability type.
They can select one of the 10 CWE entries in the vulnerability type list, including Buffer Overflow (CWE-119), Cross Site Scripting (CWE-79), SQL Injection (CWE-89), Directory Traversal (CWE-22), XML External Entity (CWE-611), Insecure Permissions (CWE-276), Incorrect Access Control (CWE-284), Integer Overflow (CWE-190), Cross Site Request Forgery (CWE-352), and Missing SSL Certification Verification (CWE-599).
If the weakness (e.g., PHP Remote File Inclusion (CWE-98), Carriage Return Line Feeds (CRLF) Injection (CWE-93)) is not in list, the reporter can select Other or Unknown, but mention the weakness in the description.

\textbf{Root cause} is an error in program design, value or condition validation, and system or environment configuration, which results in software vulnerabilities.
When submitting a new CVE request, the reporter may optionally describe the root cause in free-form natural language, as shown in Figure~\ref{fig:cvedesccomplete} and Figure~\ref{fig:cvedescmissing}.
SecurityFocus~\cite{sf} abstracts the natural language root causes of CVEs into 11 error classes: Access Validation, Atomicity, Boundary Condition, Configuration, Design, Environment, Failure to Handle Exceptional Conditions, Input Validation, Origin Validation, Race Condition, Serialization.

\textbf{Affected product} refers to [Component] in [Vendor][Product] [Version] information in the CVE description. 
It identifies software component in certain version(s) of a software product that has been affected by the CVE.
As the examples in Figure~\ref{fig:cvedesccomplete} and Figure~\ref{fig:cvedescmissing} show, affected components can be source code file, function, or executable.
When submitting a new CVE request, the reporter must provide affected product(s) and version(s), and product vendor(s).

\textbf{Attacker type} describes the way by which an attacker may exploit the CVE.
When submitting a new CVE request, the reporter may select one of the six options: 
authenticated, local, remote, physical, context dependent, or other.
Authenticated means that the attacker needs privilege or permission during the attack.
Local means that to exploit the vulnerability the attacker needs to be logged into the operating system on a local machine or a guest operating system.
Remote means that the vulnerability can be exploited through a network. The attacker may be either on the adjacent or remote network.
Physical means that the attacker needs to be located near the victim or have physical access to the vulnerable system to exploit the vulnerability, for example, touching a workstation keyboard or USB device, ``shoulder surfing'' to see a workstation's display, and touching the screen of a mobile device.
Context-dependent means that the type of access needed to exploit the vulnerability is dependent on how the vulnerable product is used. This is most often used for libraries.
Attacker type is an optional field.
That is, the reporters leave this field unspecified or as other.
But they may mention attacker type in natural language description.

\textbf{Impact} indicates what the attacker gains by exploiting this vulnerability.
When submitting a new CVE request, the reporter may select one of the five options:
code execution, information disclosure, denial of service, escalation of privileges, or other. 
Impact is an optional field, which can be left unspecified.
But the reporter generally describes the impact in natural language description.

\textbf{Attack vector} describes the method of exploitation, for example, to exploit vulnerability, someone must open a crafted JPEG file.
When submitting a new CVE request, attack vector is an optional field.
When provided, it will be only in natural language description.
We manually label attack vector descriptions into five common types: via field, arguments or parameters, via some crafted data, by executing the script, HTTP protocol correlation, call API.

\subsection{Rule-Based Extraction of CVE Aspects}
\label{sec:extractionrules}
\begin{table}
	\setlength{\abovecaptionskip}{-0cm}%
	\caption{Classes of CVE aspects and their percentages}
	\label{tab:freq}
	\scriptsize
	\begin{tabular}{ccccc}
		\toprule
		Vulnerability Type\\
		\midrule
		Cross site scripting(CWE-79)&29.5\%&SQL injection(CWE-89)&17.8\%\\
		Buffer Overflow(CWE-119)&17.1\%&Directory Traversal(CWE-32)&8.9\%\\
		Cross-site request forgery(CWE-352)&7.1\%&PHP file inclusion(CWE-98)&5.7\%\\
		Use-after-free(CWE-416)&3.2\%&Integer overflow(CWE-680)&2.6\%\\
		Untrusted search path(CWE-426)&1.7\%&Format string(CWE-134)&1.6\%\\
		CRLF injection(CWE-93)&0.6\%&XML External Entity(CWE-661)&0.3\%\\
		Others&4.0\%\\
		\toprule
		Root Cause\\
		\midrule
		Input Validation Error&51.7\%&Boundary Condition Error&24.5\%\\
		Failure to Handle Exceptional Conditions&11.7\%&Design Error&11.0\%\\
		Access Validation Error&0.7\%&Atomicity Error&0.1\%\\
		Race Condition Error&0.1\%&Serialization Error&0.1\%\\
		Configuration Error&0.1\%& Origin Validation Error&0.1\%\\
		Environment Error&0.1\%\\
		\toprule
		Attack Vector\\
		\midrule
		Via field, arguments or parameter&51.7\%&Via some crafted data&17.1\%\\
		By executing the script&14.0\%&HTTP protocol correlation&4.4\%\\
		Call API&3.3\%&Others&8.0\%\\
		\toprule
		Attacker Type\\
		\midrule
		Remote attacker&72.8\%&Local attacker&11.1\%\\
		Authenticated user&8.1\%&Context-dependent&2.9\%\\
		Physically proximate attacker&0.3\%&
		
		Others&4.7\%\\	
		\bottomrule
	\end{tabular}
	\vspace{-0mm}
\end{table}
We randomly sample 20\% of the reported CVEs per year from 1999 to 2019 (in total 24,042 CVEs examined). 
We observe that about 71\% of CVE descriptions follow the suggested templates~\cite{poj}, such as those in Figure~\ref{fig:cvedesccomplete} and Figure~\ref{fig:cvedescmissing}(a)(b).
We see the increasing trend of following the suggested templates over time due to the continual standardization effort of the CVE organization.
The rest 29\% of CVE descriptions do not follow the suggested templates, such as those in Figure~\ref{fig:cvedescmissing}(c)(d).
However, even for those non-template-following CVE descriptions, the descriptions of CVE aspects still exhibit similar patterns.
Due to the input assistance of the CVE request website, we observe commonly used phrases or their variants for vulnerability type, attacker type and impact.
We also observe a list of frequently affected products and their vendor names.

Based on our observation of the sentence regularities of CVE descriptions, we develop a set of regular expression rules\footnote{Our aspect extraction tool and all experiment data/results are available at \href{https://github.com/ pmaovr/Predicting-Missing-Aspects-of-Vulnerability-Reports}{\textcolor{blue}{https://github.com/pmaovr/Predicting-Missing-Aspects-of-Vulnerability-Reports}}} to extract CVE aspects from CVE descriptions.
First, we build a gazetteer (i.e., dictionary) of commonly used phrases for vulnerability type, root cause, impact, attacker type and attack vector, as well as a gazetteer of product and vendor names from \href{https://www.cvedetails.com/}{\textcolor{blue}{CVE Details}}.
Gazetteer based matching is a basic but very useful technique for the extraction of frequently seen concepts and entities in text~\cite{ner0,ner1}.
Phrases in the gazetteer are defined as regular expressions to match not only the extract phrases but also their variants.

We use the Stanford CoreNLP tool~\cite{stanf} to parse a CVE description sentence and obtain the POS tags of the sentence.
Next, we combine gazetteer matching and POS pattern matching to determine the candidates of certain CVE aspects.
Finally, we examine the candidates against the two officially defined sentence templates and other commonly seen sentence-level patterns.
These sentence-level patterns define the commonly seen appearance order of different aspects in a CVE description.
For example, vulnerability type or root cause often appears together with affected product at the beginning of the CVE description.
Attacker type, impact and attack vector often appear together in the form of ``allow [attacker type] to [impact] via [attack vector]'' or ''[attacker type] performs [attack vector] in order to [impact]''.
By examining the aspect candidates against the sentence patterns, we can filter out false positive aspect candidates.
For example, ``executing the script'' can be an Attack vector or Impact aspect, but putting ``executing the script'' in the sentence ``By executing the script ...'', we can determine ``executing the script'' is actually an Attack vector aspect.

\subsection{Accuracy of CVE Aspect Extraction}
\label{sec:extractionquality}


We apply our aspect extraction tool to the rest 96,081 CVEs.
We extract 41003, 15129, 92132, 89053, 73134 and 67188 instances of the vulnerability type, root cause, affected product, impact, attacker type and attack vector aspect, respectively.
Considering large numbers of instances to examine, we adopt a statistical sampling met\-hod~\cite{samp} to evaluate the accuracy of the extracted aspect instances.
Specifically, we sample and examine the minimum number MIN of data instances in order to ensure that the estimated accuracy.
MIN is determined by $n_0/(1+(n_0-1)/populationsize)$ where $n_0 = (Z^2*0.25)/e^2$, and $Z$ is the confidence level's z-score and $e$ is the error margin.
In this work, we consider 5\% error margin at 95\% confidence level.
At this setting, we examine 384 extracted instances of each aspect.
One author labels the sampled instances, and the other author validates the results.
The two authors discuss to resolve the disagreement.
The extraction accuracy is 97\%, 96\%, 96\%, 98\%, 99\% and 98\% for the aspect vulnerability type, root cause, affected product, impact, attacker type and attack vector, respectively.

\subsection{Missing and Distribution of CVE Aspects}
Based on the extracted CVE aspects (which are of high accuracy), we analyze the missing of different aspects in the CVE descriptions.
We observe different severities of information missing for different aspects.
About 43.8\% of CVEs describe specific vulnerability type.
Only about 15.2\% of CVEs describes specific root cause.
About 3\% of CVEs describe both vulnerability type and root cause.
62\% of CVEs describe attack vector, and 72\% of CVEs describe attacker type.
Almost all (over 99\%) CVEs describe affected product, and over 94\% of CVEs describe impact.
We find that about 31\% of CVE miss one aspect, 39\% missing two, and 28\% miss three or more aspects.

Table~\ref{tab:freq} lists the class distributions of the extracted aspects.
We do not include affected product and impact because they do not suffer from significant information missing.
We can see that the class distribution of an aspect is in general imbalance, which has several frequent classes and many less frequent classes in the long tail.
We group the classes with the frequencies $<0.1$ as Others.

\section{Neural-Network Based Prediction of CVE Missing Aspects}
\label{sec:approach}

\begin{figure*}
	\centering
	\setlength{\abovecaptionskip}{0.8pt}
	\setlength{\belowcaptionskip}{1.0pt}
	\includegraphics[scale=0.35]{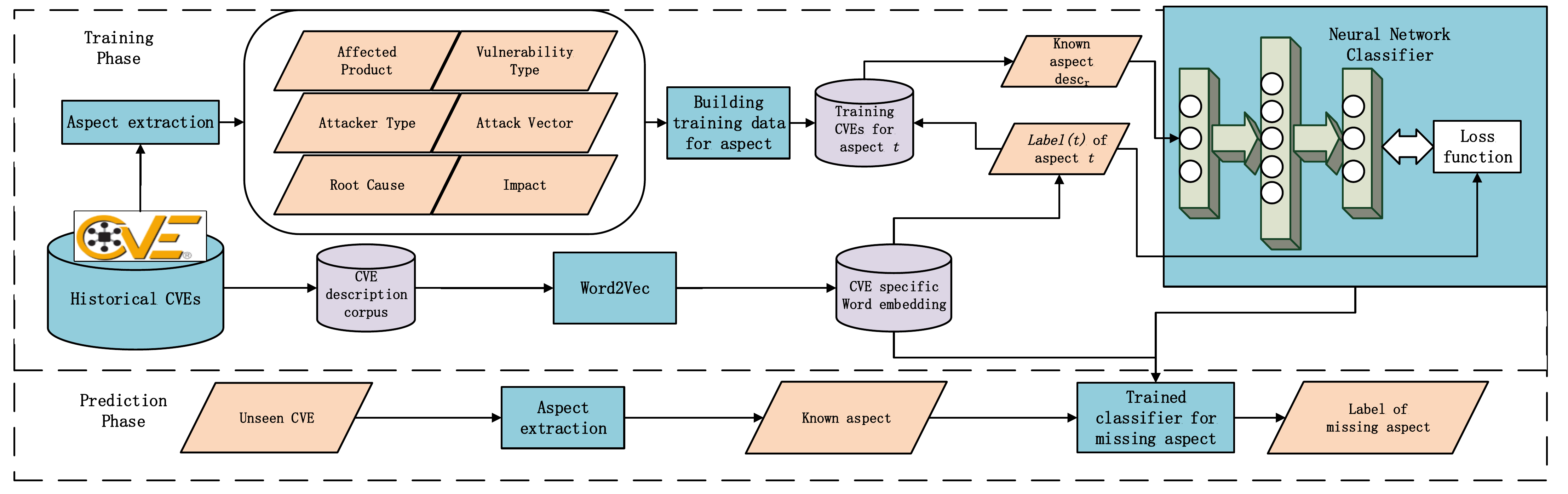}
	\caption{Approach Overview}
	\label{fig:approachoverview}
	\vspace{-5mm}
\end{figure*}

\begin{figure}
	\centering
	\setlength{\abovecaptionskip}{0pt}
	\setlength{\belowcaptionskip}{0pt}
	\includegraphics[scale=0.7]{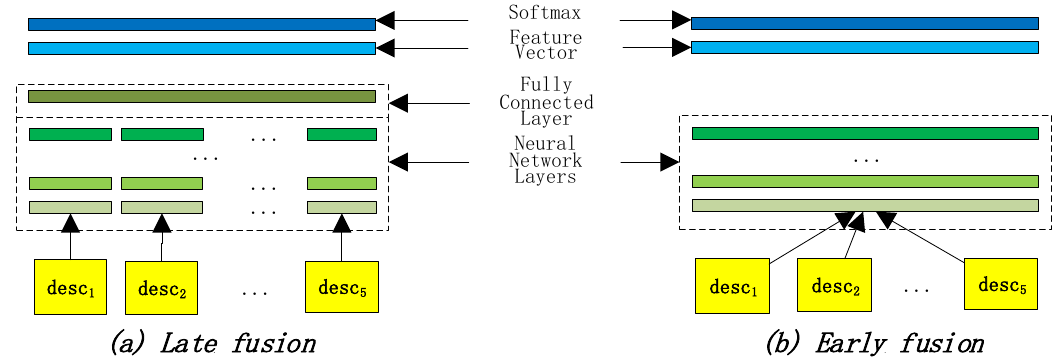}
	\caption{Two model architectures}
	\label{fig:modelarchitecture}
	\vspace{-5mm}
\end{figure}
Motivated by the missing of key aspects in CVE descriptions, we design a neural-network based approach for predicting the missing aspects based on the known aspects in a CVE description.
We first give an overview of our approach (Section~\ref{sec:approachoverview}) and then describe the design of neural-network classifier (Section~\ref{sec:input}-Section~\ref{sec:output}).

\subsection{Approach Overview}
\label{sec:approachoverview}

We formulate the prediction task as a multi-class classification problem for each aspect.
Considering the severity of the information missing (see Section~\ref{sec:extractionquality}), we predict four aspects: vulnerability type, root cause, attacker type and attack vector.
Each aspect has a corresponding multi-class classifier, and the class labels for each aspect-specific classifier are summarized in Table~\ref{tab:freq}.

As shown in Figure~\ref{fig:approachoverview}, our approach consists of a training phase and a prediction phrase.
Training phase uses the historical CVE descriptions to train aspect-specific neural-network classifiers.
It first uses the aspect extraction method in Section~\ref{sec:extractionrules} to extract six CVE aspects from the historical CVE descriptions.
Then, we prepare the training data for each aspect $t$ to be predicted.
For each CVE that contains the aspect $t$, a training instance is created with the class label of $t$ (denoted as $label(t)$) as the expected output and the description of rest of aspects $r \in R$ ($1 \leq |R| \leq 5$) (denoted as $desc(r)$) as the input.
From such training data, the neural-network classifier is trained to extract syntactic and semantic features from the input aspect descriptions and capture the intrinsic correlations between these input features and the output class label.

At the prediction phrase, given an unseen CVE description, we first extract the CVE aspects present in the description.
For each missing aspect, the trained aspect-specific classifier takes as input the aspects present in the description and predicts as output the most likely class label of the missing aspect. 

The neural network classifier consists of three layers:
an input layer that represents the input text in a vector representation (e.g., word embedding) (Section~\ref{sec:input});
a neural-network feature extractor that extracts syntactic and semantic features from the input text (Section~\ref{sec:nn});
and an output classifier that makes the prediction based on the extracted features (Section~\ref{sec:output}).
Next, we describe the design of these three layers in details.
\vspace{-0.2cm}
\subsection{Input Text and Representation}
\label{sec:input}

The raw input to the classifier is the textual description of CVE aspects, such as those sentence fragments highlighted in Figure~\ref{fig:cvedesccomplete} and Figure~\ref{fig:cvedescmissing}.
In this work, we consider three formats of raw input text:
1) the sequence of separate aspect descriptions in the original appearance order (denoted as \textit{i-ao});
2) the sequence of separate aspect descriptions in a random order (denoted as \textit{i-ar});
3) the original CVE description containing all input aspects (denoted as \textit{i-fu}).
\textit{i-ar} allows us to investigate the impact of the appearance order of CVE aspects, and \textit{i-fu} allows us to investigate the impact of additional sentence parts in the original CVE description.
Additional parts refer mostly to preposition, pronoun and/or determiner that connect separate aspect descriptions into a more complete sentence.

Words are discrete symbols.
As the input to the neural network, they must be represented in vectors.
In this work, we use word embeddings to represent words in a vector space, because many studies~\cite{we1, we2, we3} have shown that word embeddings can capture rich syntactic and semantic features of words in a low-dimensional vector.
We consider both general word embeddings (denoted as $we_g$) and domain-specific word embeddings (denoted as $we_d$).
For general word embeddings, we use word embeddings pre-trained on the corpus of Google News text directly
obtained from the official word2vec website~\cite{googlenews}.
Google News word embedding is learned by continuous skip-gram model~\cite{we4}.
For domain-specific word embeddings, we compile two corpora: one from CVE descriptions and the other from the vulnerability report in SecurityFocus.
We set the vocabulary size at 50,000 and learns domain-specific word embeddings on these two corpora separately using continuous skip-gram model (the Python implementation in Gensim~\cite{we5}).
The output of word embedding learning is a dictionary of words, each of which has a $d$-dimensional vector.
In this work, we set $d$ at 300 as in existing studies~\cite{han2017learning,gong2019joint,xiao2019embedding}.

Let the input text be a sequence of $N$ words $w_i$ ($1 \leq i \leq N$).
This input text is represented as a $N \times d$ matrix: $i$=$v(w_1) \oplus v(w_2)  \oplus ...\oplus v(w_N)$, where $\oplus$ is vector concatenation and $v(w_i)$ returns the word embedding of the word $w_i$ in the dictionary.
We randomly initialize corresponding word vectors to deal with these Out-of-Vocabulary (OOV) words~\cite{oov2}.
Different CVEs may have different numbers of known CVE aspects.
To keep the model architecture consistent, we set an input aspect as an empty string if the CVE does not contain this input aspect. 

\subsection{Neural Network Feature Extractors}
\label{sec:nn}

A neural-network feature extractor takes as input the matrix $i$ from the input layer and computes as output a feature vector $o$ which is fed into the output Softmax classifier for prediction.

\subsubsection{Model Architecture}
\label{sec:model}

As our input consists of separate CVE aspects, we design two model architectures to investigate the effective mechanism for incorporating CVE aspects and capturing their intrinsic correlations: early fusion versus late fusion.
As shown in Figure~\ref{fig:modelarchitecture}, early fusion architecture first concatenates the input matrix of each aspect into one input matrix, which is fed into a single neural network to extract and fuse features from different aspects. 
In contrast, late fusion architecture feeds the input matrix of each aspect into a neural network separately and then fuse the output feature vector of the separate networks by a fully connected layer. 
All neural networks share the same network configuration, but they will learn different weights in different architectures.
Both early fusion and late fusion are applicable to the input formats \textit{i-ao} and \textit{i-ar}.
But only early fusion is applicable to the input format \textit{i-fu}, because \textit{i-fu} merges the input CVE aspects into a whole sentence.

\subsubsection{Backbone Network}
\begin{figure}
	\setlength{\abovecaptionskip}{0mm}
	\setlength{\belowcaptionskip}{0mm}
	\centering
	
	\includegraphics[scale=0.23]{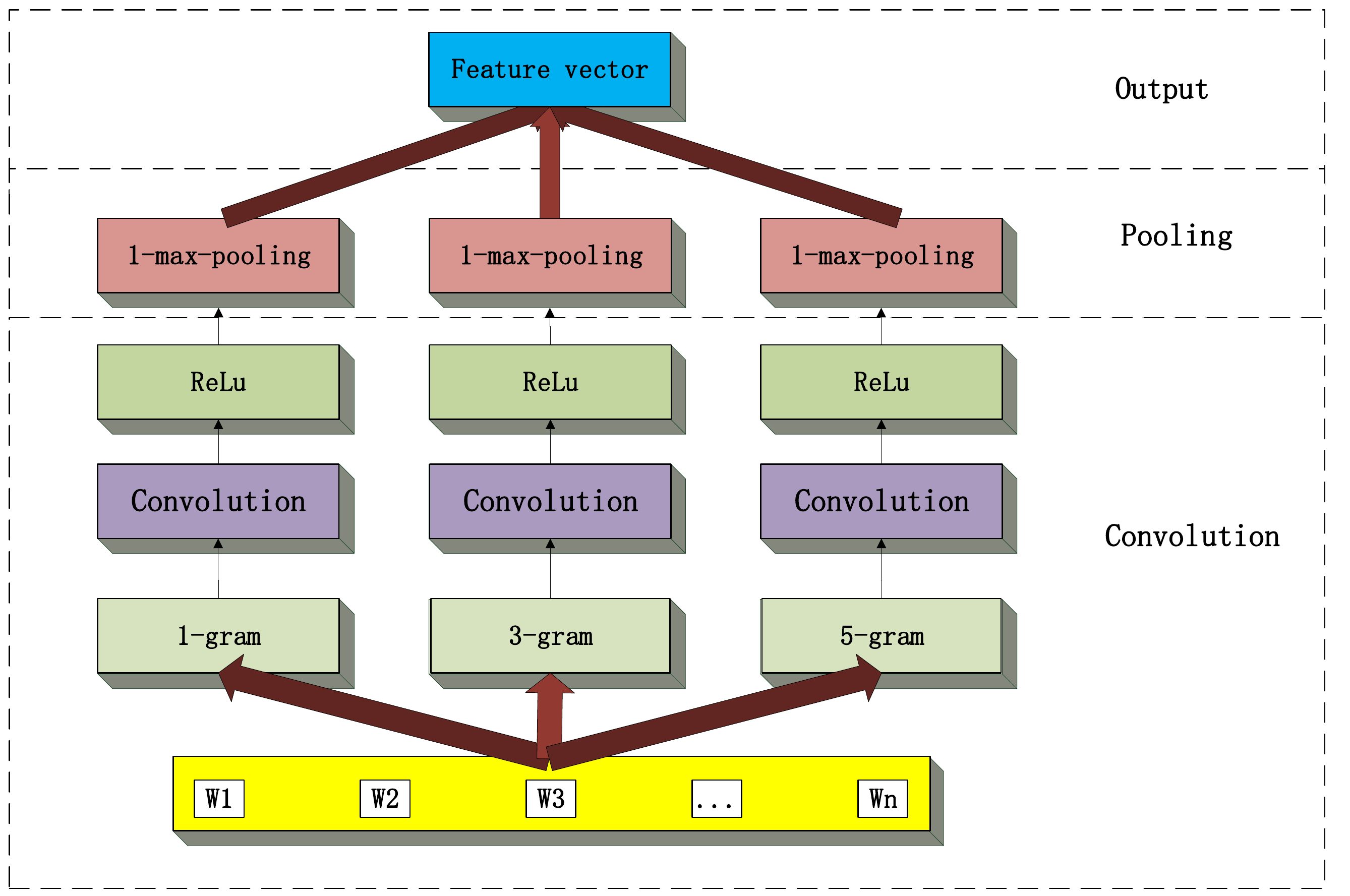}
	
	\caption{1-layer CNN}
	\label{fig:cnn}
\vspace{-5mm}
\end{figure}

We consider two popular neural netwo\-rks used for text classification in the literature~\cite{Hassan2018Convolutional,Gurulingappa2012Development,Howard2018Universal}:
Convolutional Neural Network (CNN) and Bi-directional Long-Short Term Memory (BiLSTM).
CNN is good at capturing important words and phrases in the text, while BiLSTM is good at capturing longer-range dependencies in text.
We design various variants of these two networks.
First, we consider shallow (1-layer) versus deep ($>1$ layer) neural network.
Second, we consider adding  attention layer on top of LSTM layer to weight word importance.
As such, we obtain six neural network models for feature extraction:

\textbf{1-layer CNN:}
Figure~\ref{fig:cnn} shows 1-layer CNN model, which consists of one convolution layer and a 1-max pooling layer.
The convolution layer applies $M$ filters to the input matrix $i$ of word embeddings.
A filter is a $h \times w$ matrix whose values will be learned during model training.
$h$ is called height or window size.
It refers to the number of consecutive words (e.g., $n$-gram) the filters apply to.
In this work, we use three different window sizes $h$=1, 3, 5. 
That is, the filters extract features from 1-grams, 3-grams and 5-grams respectively.
The width $w$ of filters is set to the dimensionality $d$ of word embeddings because the filters should retain the integrity of the input word embeddings.
For each word window, a filter computes a real value, which is fed into the non-linear activation function ReLU~\cite{relu}: ReLU(x)=max(0,x).
A filter scans the input word sequence (zero-padding at both ends to allow the filter to extract features from the beginning and the end of the input sentence) with stride=1.
This type of filters is also known as 1-d convolution as it scans the input matrix along only one dimension (i.e., the word sequence).
After scanning the whole input sequence, a filter generates a feature map of the input sequence length.
A 1-max pooling is applied to this feature map to obtain the most significant feature for this filter.
Due to the use of 1-max pooling, we do not need attention layer to weight word/phrase importance. 
For each window size, we use $M$=128 filters to learn complementary features from the same word windows.
That is, 1-layer CNN produces a 128-dimensional feature vector for each window size.
The feature vectors of all windows sizes are concatenated into an output feature vector for the classifier.

\begin{figure}
	\setlength{\abovecaptionskip}{0mm}
	\setlength{\belowcaptionskip}{0mm}
	\centering
	\includegraphics[scale=0.25]{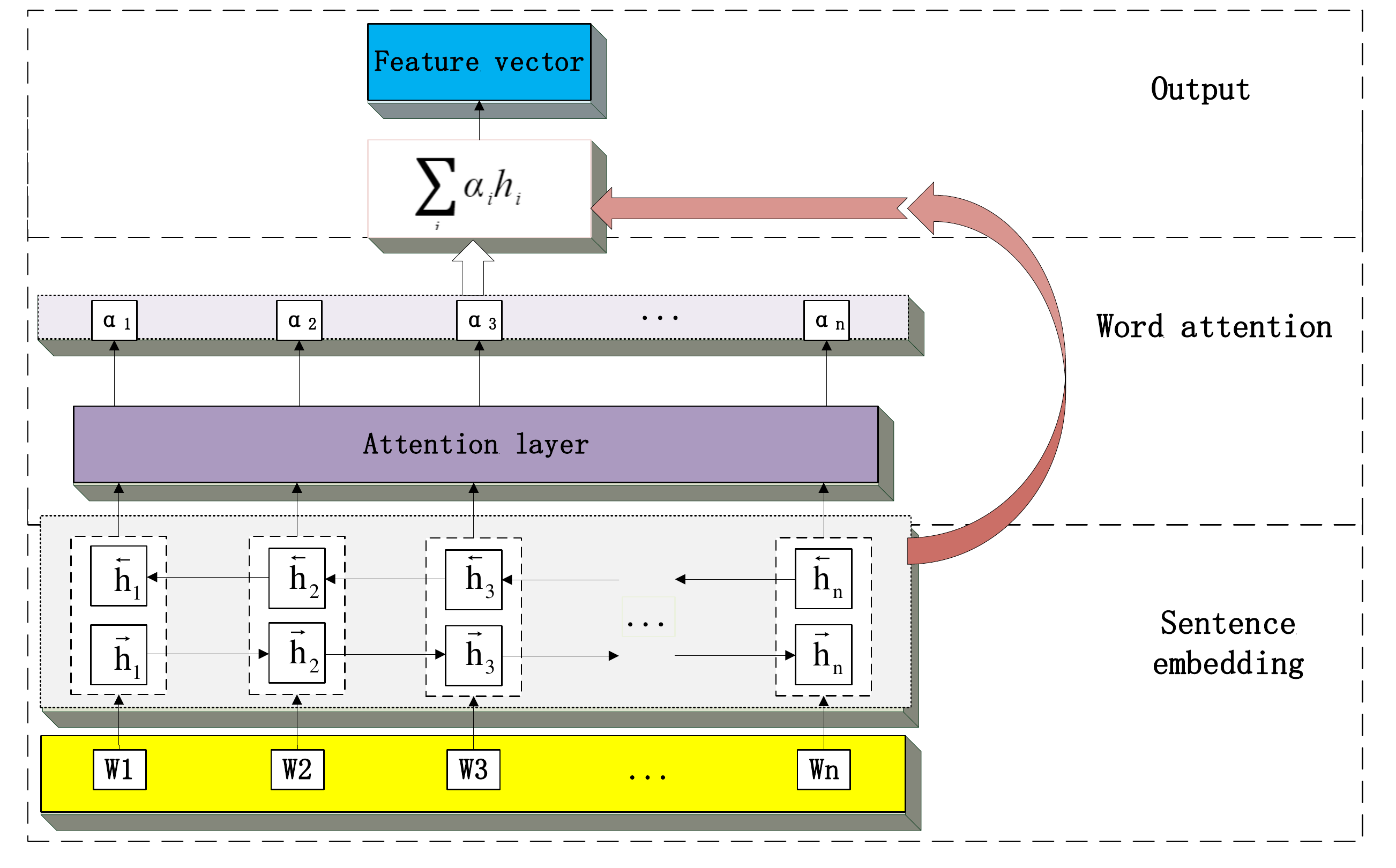}
	
	\caption{1-layer Bi-LSTM with attention}
	\label{fig:lstm}
\vspace{-5mm}
\end{figure}

\textbf{2-layers CNN:}
This is a deeper variant of the 1-layer CNN.
The first CNN layer is the same as 1-layer CNN.
The three feature vectors by the first CNN  layer are fed into one more layer of convolution and 1-max-pooling.
For each input $128$-dimensional feature vector, the second layer also uses $M$ ($M=128$) filters with the window size $h=3$.
After the convolution, ReLu activation and pooling, the second CNN layer outputs three $128$-dimensional feature vectors which are concatenated into an output feature vector for the classifier.
 
\textbf{1-layer BiLSTM:}
Figure~\ref{fig:lstm} shows a 1-layer BiLSTM model.
A BiLSTM model consists of a forward LSTM $\overrightarrow{lstm}_f$ network that reads the input from $w_1$ to $w_N$, and a backward LSTM $\overleftarrow{lstm}_b$ that reads from $w_N$ to $w_1$:
$\overrightarrow{h}_{i} = \overrightarrow{lstm}_f(w_i), i\in\left[1,n\right]$,
$\overleftarrow{h}_{i} = \overleftarrow{lstm}_b(w_i),\\ i\in\left[n,1\right]$.
$\overrightarrow{h}_{i}$ and $\overleftarrow{h}_{i}$ are forward and backward hidden vectors for word $w_i$.
Both forward and backward LSTM have 192 LSTM cells.
So $\overrightarrow{h}_{i}$ and $\overleftarrow{h}_{i}$ are 192-dimensional vectors.
The parameters of $\overrightarrow{lstm}_f$ and $\overleftarrow{lstm}_f$ will be learned during model training.
We obtain the hidden vector $h_i$ for $w_i$ by concatenating $\overrightarrow{h}_{i}$ and $\overleftarrow{h}_{i}$, i.e., $h_i = \overrightarrow{h}_{i}\oplus\overleftarrow{h}_{i}$ ($h_i$ is 384-dimensional vector).
$h_i$ is the output vector of the Bi-directional LSTM for the input word $w_i$, which encodes both the preceding and succeeding sentence context centered around $w_i$.
We concatenate the last hidden vectors $\overrightarrow{h}_{N}$ and $\overleftarrow{h}_{1}$ into an output feature vector for the classifier.

\textbf{2-layers BiLSTM:}
This is a deeper variant of 1-layer BiLSTM.
The first layer is a 1-layer BiLSTM.
The output $h_i$ of the first BiLSTM layer is passed as input to the second BiLSTM layer which also uses 192 LSTM cells to encode input vector sequence.
The last hidden vectors $\overrightarrow{h}_N$ and $\overleftarrow{h}_1$ by the second layer are concatenated into an output feature vector for the classifier.

\textbf{Attention layer for BiLSTM:}
For the BiLSTM network, we can add an attention layer on top of the last BiLSTM layer to weight the importance of words.
The attention layer computes the word attention weight $\alpha_i = \exp\left(\left(u_i\right)^\mathsf{T} u_w\right)/\sum_{i}\exp\left(\left(u_i\right)^\mathsf{T} u_w\right)$ where $u_i = \tanh\left(W_w h_i + b_w\right)$.
It takes the output $h_i$ of the BiLSTM layer as input.
$h_i$ is first fed through a one-layer feed-forward neural network to get $u_i$ as a hidden representation of $h_i$.
$W_w$ and $b_w$ are learnable parameters of the attention layer.
Then, we measure the importance of the word $w_i$ as the similarity of $u_i$ with a word-level context vector $u_w$ and obtain a normalized importance weight $\alpha_i$ through a softmax function.
The vector $u_w$ is randomly initialized and learned during the training process.
Finally, we compute the output feature vector for the classifier as a weighted sum of the Bi-LSTM output vector of each word in the input sequence multiplied by the word attention weight, i.e., $\sum_{i=1}^n\alpha_i h_i$.

\vspace{-2mm}
\subsection{Predicting Missing Aspects of CVEs}
\label{sec:output}

We use a softmax classifier as the output classifer. 
Given the output feature vector $F$ by the neural-network feature extractor, a softmax classifier predicts the probability distribution $\hat{y}$ over the $m$ class labels of a particular CVE aspect, i.e., $\hat{y}=softmax\left(WF + b\right)$, where $\hat{y} \in \mathcal{R}^{m}$ is the vector of prediction probabilities over the $m$ class labels, and $W$ and $b$ are the learnable parameters of the classifier.
The learnable parameters of the neural-network feature extractor and the classifier are trained to minimize the cross-entropy loss of the predicted labels and the ground-truth labels: $L\left(\hat{y},y\right)=-\sum_{i=1}^K\sum_{j=1}^m y_{ij}\log\left(\hat{y}_{ij}\right)$ where $K$ denotes the number of training samples.
$y_{ij}$ is the ground-truth label of the $j$th class (1 for the ground-truth class, otherwise 0) for the $i$th training example, and $\hat{y}_{ij}$ is the predicted probability of the $j$th class for the $i$th training example.
The loss gradient is back-propagated to update the learnable parameters of the classifier and the neural-network feature extractor.



\section{Experiments}

We conduct a series of experiments to investigate the following four research questions:

\begin{itemize} [leftmargin=*]
	\item \textbf{RQ1 Design of neural network classifier}: How do different input formats, word embeddings, model architectures and neural network designs affect the prediction performance?
	
	\item \textbf{RQ2 Overall performance}: How well can our approach predict different CVE aspects?
	
	\item \textbf{RQ3 Size of training data}: How does the size of training data affect the prediction performance? 
	
	\item \textbf{RQ4 Ablation study}: What is the most and least prominent aspect or aspect combination for predicting a specific aspect?
\end{itemize}
\vspace{-1mm}
\subsection{Experiment Setup}

We describe the CVE dataset used in our experiments, our model training setting, and the performance evaluation metrics.

\subsubsection{CVE Dataset}
\label{sec:dataset}


CVE list can be downloaded from the official CVE website~\cite{cve}.
In this work, we download the CVE list that contains 120,103 CVEs from January 1999 to October 2019.
We use the aspect extraction method in Section~\ref{sec:extractionrules} to extract CVE aspects from these CVEs.
Our evaluation in Section~\ref{sec:extractionquality} confirms the high accuracy of the extracted CVE aspects.
To evaluate our prediction method, we collect 51,803 CVEs whose descriptions contain at least four aspects.
As almost all CVEs contain affected product and impact aspects (see Section~\ref{sec:extractionquality}), this means the CVEs in the dataset contain at least two of the other four aspects.
This guarantees that we have sufficient data to study aspect fusion and ablation.
For each to-be-predicted aspect (vulnerability type, root cause, attacker type or attack vector), we build an aspect-specific dataset for classifier training and testing from these 51,803 CVEs according to the method in Section~\ref{sec:approachoverview}.
Table~\ref{tab:dataset} summarizes the information of the four aspect-specific datasets.
All experiment datasets and results are available at  \href{https://github.com/ pmaovr/Predicting-Missing-Aspects-of-Vulnerability-Reports}{\textcolor{blue}{our Github repository}}.
For a specific to-be-predicted aspect (e.g., vulnerability type), the other three aspects (e.g., root cause, attacker type or attack vector) of the CVEs used as model input may also be missing, except affected product and impact.
This reflects the realistic situation of CVE data.
That is, when predicting a missing aspect, it is often not all other aspects available as model input.
\begin{table}
\setlength{\abovecaptionskip}{-0.05cm}%
	\setlength{\belowcaptionskip}{0cm}%
	\caption{Aspect-specific CVE datasets}
	\label{tab:dataset}
	\scriptsize
	\setlength{\tabcolsep}{0.9mm}{
	\begin{tabular}{ccccc}
		\toprule
		To-be-predicted aspect&Vulnerability type&Root cause&Attack vector&Attacker type\\

		Size&36,103&20,813&35,289&43,330\\
\midrule
		\% with vulnerability type&100\%&61\%&63\%&66\%\\
		\% with root cause&38\%&100\%&39\%&37\%\\
		\% with attack vector&75\%&72\%&100\%&78\%\\
		\% with attacker type&86\%&86\%&92\%&100\%\\
				
		\bottomrule
	\end{tabular}}
\vspace{-6mm}
\end{table}

\subsubsection{Model Training}

We implement the proposed neural network classifier in Tensorflow~\cite{Abadi2016TensorFlow}.
Each to-be-predicted aspect has its own classifier. 
As discussed in Section~\ref{sec:approach}, we have different choices for input format, word embedding, model architecture and network design when implementing a classifier.
All the classifiers are trained in the same setting.
Specifically, we train each model for 256 iterations with a batch size of 128, set learning rate at 0.001, and use Adam~\cite{Kingma2014Adam} as the optimizer.
All experiments run on an NVIDIA Tesla M40 GPU machine.

\subsubsection{Evaluation Metrics}
\label{sec:metrics}

The multi-class classification results can be represented in a $m \times m$ confusion matrix $M$, where $m$ is the number of class labels (see Table~\ref{tab:freq}).
We use Precision, Recall and F1-score to evaluate the effectiveness of multi-class classification~\cite{mt1,mt2,mt3,mt4}.
Precision for a label $L_j$ of an aspect $A$ represents the proportion of the CVEs whose missing aspect $A$ is correctly predicted as $L_j$ among all CVEs whose missing aspect $A$ is predicted as $L_j$.
Recall for a label $L_i$ of an aspect $A$ is the proportion of the CVEs whose missing aspect $A$ is correctly predicted as $L_i$ compared with the number of ground-truth CVEs whose missing aspect $A$ is actually $L_i$.
F-score is the harmonic average of the precision and recall.
The overall performance of a classifier is the weighted average of the evaluation metrics of each class label. 
Since F1-score conveys the balance between the precision and the recall, we use F1-score as the main evaluation metric in the discussion.
\subsection{Design of Neural Network Classifier (RQ1)}

\noindent
\textbf{Motivation:}
The design of our neural network classifier considers three input formats (i-ao: separate CVE aspects in the original order appearing in CVE descriptions, i-ar: separate CVE aspects in random order, and i-fu: original CVE descriptions), three word embeddings (Google News, SecurityFocus versus CVE-specific), two model architecture (early aspect fusion versus late aspect fusion), and six specific neural network design (CNN versus BiLSTM, 1-layer versus 2-layer, BiLSTM with/without attention layer).
We want to investigate the impact of these design options on the prediction performance and identify the most effective design of neural network classifier.

\vspace{0.5mm}
\noindent
\textbf{Approach:}
We conduct four experiments to evaluate the impact of input format, word embedding, model architecture and network design respectively.
For the experiments on one dimension, we use the most effective options for the other three dimensions.
Specifically, for input format experiments, we use CVE-specific word embeddings, early fusion architecture and 1-layer CNN.
For word embedding experiments, we use separate CVE aspects in original order, early fusion architecture and 1-layer CNN.
For model architecture experiments, we use separate CVE aspects in original order, CVE-specific word embeddings and 1-layer CNN.
For network design experiments, we use separate CVE aspects in original order, CVE-specific word embeddings and early fusion architecture.
This experiment setting helps to reduce the large number of experiments by the full Cartesian product combination of the design options, and also facilitate the analysis of each design dimension while fixing the other three dimensions.
To ensure the reliability of our experiments, we perform 10-fold cross validation in all the experiments.
For each fold, we use 80\%, 10\% and 10\% of data for model training, hyperparameter optimization and testing respectively.
We conduct Wilcoxon signed-rank test~\cite{woolson2007wilcoxon} on F1-score between different experiment settings.
$p$-value $<0.05$ is considered statistically significant (marked by * in the results tables).

\begin{table}
\setlength{\abovecaptionskip}{-0.05cm}%

	\centering
	\scriptsize
	
	\caption{\label{tab:inputformats} Impact of input formats}
	
\setlength{\tabcolsep}{0.9mm}{
	\begin{tabular}{cccccccc}

		\toprule
		& & & Vulnerability Type& Root Casue& Attack Vector &Attacker Type\\
		\midrule	
	
		\multirow{3}{*}{Precision}
		&&i-ao & 0.945& 0.779 & 0.708 & 0.884\\
		&&i-ar& 0.943& 0.746& 0.701 & 0.882\\
		&&i-fu  & 0.946 & 0.783 & 0.703&0.888 \\
		\midrule

		\multirow{3}{*}{Recall}
		&&i-ao& 0.945& 0.793& 0.716 & 0.897\\
		&&i-ar& 0.945 & 0.770 & 0.710 &0.892\\
		&&i-fu& 0.946& 0.796 & 0.717 &0.899\\
		\midrule	
	
		\multirow{3}{*}{F1}
		&&i-ao & 0.943& 0.780& 0.704 &0.885\\
		&&i-ar & 0.943 & 0.745 & 0.699 &0.880\\
		&&i-fu& 0.946& 0.788 & 0.706 &0.889\\
		\bottomrule
	\end{tabular}
}\vspace{-3mm}
\end{table}

\begin{table}
\setlength{\abovecaptionskip}{-0.05cm}%
	\centering
	\scriptsize
	\caption{\label{tab:wordembedding}Impact of word embeddings}

\setlength{\tabcolsep}{0.9mm}{
	\begin{tabular}{llccccc}
		\toprule
		& & & Vulnerability Type& Root Casue& Attack Vector &Attacker Type\\
		\midrule	
	
		\multirow{3}{*}{Precision}&&CVE& 0.946  & 0.783& 0.703&0.888 \\
		&&SecurityFocus& 0.942  & 0.779& 0.701& 0.883\\
		&&*Google news& 0.932 & 0.759& 0.687 & 0.869\\
		
		\midrule

		\multirow{3}{*}{Recall}&&CVE& 0.946 & 0.796& 0.717 &0.899\\
		&&SecurityFocus& 0.944 & 0.792& 0.716  &0.894\\
		&&*Google news& 0.935  & 0.784& 0.688& 0.885\\
		
		\midrule	
	
		\multirow{3}{*}{F1}&&CVE& 0.946 & 0.788& 0.706 &0.889\\
		&&SecurityFocus& 0.942 & 0.783& 0.703  &0.883\\
		&&*Google news& 0.933& 0.761 & 0.687 & 0.871\\
		\bottomrule
	\end{tabular}
}	\vspace{-3mm}
\end{table}

\begin{table}
\setlength{\abovecaptionskip}{-0.05cm}%
	\centering
	\scriptsize
	
	\caption{\label{tab:modelarchitecture}Impact of model architectures}

	\setlength{\tabcolsep}{0.9mm}{
		\begin{tabular}{llccccc}
			\toprule
			& & & Vulnerability Type& Root Casue& Attack Vector &Attacker Type\\
			\midrule	
			
			\multirow{3}{*}{Precision}&&Early Fusion& 0.946 & 0.783& 0.703 &0.888 \\
			&&*Late Fusion& 0.921 & 0.751& 0.671 & 0.844\\
			
			\midrule

			\multirow{3}{*}{Recall}&&Early Fusion& 0.946& 0.796 & 0.717 &0.899\\
			&&*Late Fusion& 0.927 & 0.770& 0.680  &0.872\\
			
			\midrule	
			
			\multirow{3}{*}{F1}&&Early Fusion& 0.946 & 0.788& 0.706 &0.889\\
			&&*Late Fusion & 0.923  & 0.755 & 0.669&0.850\\
			\bottomrule
		\end{tabular}
	}\vspace{-5mm}
\end{table}

\begin{table}
	\centering
	\scriptsize
	\setlength{\abovecaptionskip}{0cm}%
	\setlength{\belowcaptionskip}{0cm}
	\caption{\label{tab:detail} Impact of neural network designs}
	
	\setlength{\tabcolsep}{0.5mm}{\begin{tabular}{cccccccc}
		\toprule
		& & & Vul-Type& Root Casue& Attack Vector &Attacker Type\\
		\midrule
		\multirow{7}{*}{Precision}&&1-L CNN  & 0.946& 0.783 & 0.703 &0.888 \\
		&&2-L CNN  & 0.933 & 0.765& 0.673 & 0.852\\
		
		&& 1-L BiLSTM      &  0.939 & 0.761& 0.682& 0.867\\
		&&2-L BiLSTM               & 0.939 & 0.770&0.688 & 0.870\\	
		&&1-L BiLSTM+Attention      &  0.941 & 0.769& 0.690& 0.873\\
		&&2-L BiLSTM+Attention   &  0.943 & 0.778& 0.692&0.876\\
				
		\midrule
		\multirow{7}{*}{Recall} &&1-L CNN             & 0.946 & 0.796& 0.717 &0.899\\
		&&2-L CNN             & 0.935  & 0.775& 0.701& 0.878 \\
		
		&&1-L BiLSTM      &  0.938 & 0.778&0.706& 0.882\\
		&&2-L BiLSTM               & 0.941 & 0.780&0.703 & 0.883\\
		&&1-L BiLSTM+Attention             & 0.943&0.778 &0.713& 0.887 \\
		&&2-L BiLSTM+Attention                & 0.945& 0.792 &0.714 & 0.889\\

		\midrule
		\multirow{7}{*}{F1-Measure}  &&1-L CNN             & 0.946 & 0.788& 0.706&0.889\\
		&&2-L CNN             & 0.932 & 0.768 & 0.677& 0.859 \\
		
		&&1-L BiLSTM      &  0.938 & 0.765& 0.684& 0.871\\
		&&2-L BiLSTM               & 0.940 & 0.770&0.683 & 0.874\\

		&&1-L BiLSTM+Attention            & 0.940& 0.770 & 0.692 & 0.873\\
		&&2-L BiLSTM+Attention             & 0.943& 0.778 & 0.694&0.878\\
		  
		\bottomrule
	\end{tabular}}
	\vspace{-3mm}
\end{table}
\subsubsection{Input Formats}

Table~\ref{tab:inputformats} presents the results.
We can see that the three input formats do not statistically significantly affect the prediction of the four CVE aspects, with only 0.002-0.009 difference in F1 across the three input formats.
The only exception is the prediction of root cause by separate aspects in random order (i-ar), but the difference in F1 is not very large either.
This suggests that the phrase-level information in the CVE aspects alone can support reliable prediction. 
The presence or absence of the additional information (mostly prepositions, pronouns, determiner) that connect CVE aspects in the original CVE descriptions does not significantly affect the prediction.
Furthermore, the prediction is not sensitive to the appearance order of different aspects in the CVE descriptions.
Therefore, we use separate aspects in the original appearance order (i-ao) as the default option.

\subsubsection{General versus Domain-Specific Word Embeddings}

Table~\ref{tab:wordembedding} shows that domain-specific word embeddings (CVE and SecurityFocus) support more accurate prediction in all four aspects than general word embeddings (Google News).
The differences are statistically significant in F1.
However, the two domain-specific word embeddings have marginal differences.
This result can be attributed to two reasons.
First, domain-specific word embeddings learn meaningful embeddings for domain-specific terms (e.g., CRLF, XSS, DoS), which may be regarded as out-of-vocabulary words in general word embeddings.
Second, domain-specific corpus allows the learning of ``purer'' word embeddings highly relevant to a particular domain, while the word embeddings learned from general text may embed some unnecessary  ``noise'' irrelevant to the particular domain.
In this work, we use CVE-specific word embeddings as the default option.

\subsubsection{Early Fusion versus Late Fusion}
\label{sec:aspectfusion}
Table~\ref{tab:modelarchitecture} shows that early fusion architecture performs better than late fusion architecture (statistically significant for all evaluation metrics).
This suggests that using a single network to extract and fuse features directly from all input CVE aspects is much more effective than extracting features from each CVE aspect separately and only fusing the features of different aspects at the end.
Therefore, we use early aspect fusion as the default option.

\subsubsection{Neural Network Variants}

Table~\ref{tab:detail} present our experimental results on the six variants of neural network feature extractor.
We can see that 1-layer CNN outperforms the other five variants.
So we use 1-layer CNN as the baseline to analyze the performance of the other five variants.
Compared with 1-layer CNN, 2-layer CNN has worse but statistically non-significant performance for predicting vulnerability type and attacker vector, but has statistically significant worse performance for predicting root cause and attacker type.
This suggests that deeper CNN is less appropriate than 1-layer CNN in our text classification task.
The performance of the four BiLSTM networks are very close.
Neither deeper BiLSTM nor attention mechanism statistically significantly improve the prediction performance.
1-layer CNN has better performance than 2-layer BiLSTM with attention (the overall best BiLSTM performer).
Although the performance differences are not large, the differences in F1 are statistically significant for predicting all four CVE aspects.
This result suggests that using CNN to extract important words/phrase features fits better for our text classification task than using LSTM to learn long-range sentence features.

\vspace{1mm}
\noindent\fbox{\begin{minipage}{8.4cm} \emph{According to our experiments on input formats, word embeddings, model architectures and network designs, the most effective design of the classifier takes as input separate CVE aspects in original order, uses CVE-specific word embeddings to represent input text, and adopts early-fusion architecture and 1-layer CNN as feature extractor.} \end{minipage}}\\

\subsection{Overall Performance (RQ2)}

\noindent
\textbf{Motivation:}
In this work, we predict four aspects: vulnerability type, root cause, attacker type and attack vector, which suffer from different severities of information missing in CVE descriptions (see Section~\ref{sec:extractionquality}).
Although vulnerability type is a compulsory aspect to enter when submitting the new CVE request, 56\% of CVE descriptions do not provide vulnerability type by leaving it as Other or Unknown.
Root cause, attacker type and attack vector are optional.
About 85\%, 28\% and 38\% of CVEs do not provide root cause, attacker type and attack vector, respectively.
We want to see how well our approach can augment these missing aspects for CVEs.

\vspace{0.5mm}
\noindent
\textbf{Approach:} 
We study the prediction performance of the most effective classifier design identified in RQ1.

\vspace{0.5mm}
\noindent
\textbf{Results:}
The precision, recall and F1 results can be found in the 1-layer CNN rows in Table~\ref{tab:detail}.
Our approach achieves 0.946 in F1 for predicting vulnerability type.
It also achieves very high F1 (0.889) for predicting attacker type.
As discussed in Section~\ref{sec:preliminary}, vulnerability type and attacker type are clearly defined categorical values.
Each class label has a distinct semantic.
This makes it easier to make accurate prediction on vulnerability type and attacker type.

In contrast, the prediction accuracy for root cause is worse. 
Compared with the distinct vulnerability types and attacker types, the error classes of root cause are relatively less distinguishable, for example, configuration versus environment, boundary condition versus input validation, access validation versus origin validation.
The fuzziness of these error classes makes it more challenging to accurately predict root cause.
Another challenge for predicting root cause is that there are six classes with only a very small number of instances, for example, atomicity error, race condition error (see Table~\ref{tab:freq}).
This is referred to as few-shot learning~\cite{wang2019few} which affects the model performance.

Our approach does not perform very well for attack vector (only 0.706 in F1).
This is because attack vector is often highly related to the information of some specific software products or vulnerabilities.
This is make it difficult to generalize the prediction model from training CVEs to unseen CVEs.

\vspace{1mm}
\noindent\fbox{\begin{minipage}{8.4cm} \emph{Our approach can accurately augment the missing vulnerability types and attacker types. However, it has limitations to predict fuzzy root causes and product- or vulnerability-specific attack vectors.} \end{minipage}}\\

\begin{figure}
	\vspace{-1mm}
	\centering
	\setlength{\abovecaptionskip}{0mm}
	\setlength{\belowcaptionskip}{1mm}
	\includegraphics[scale=0.5]{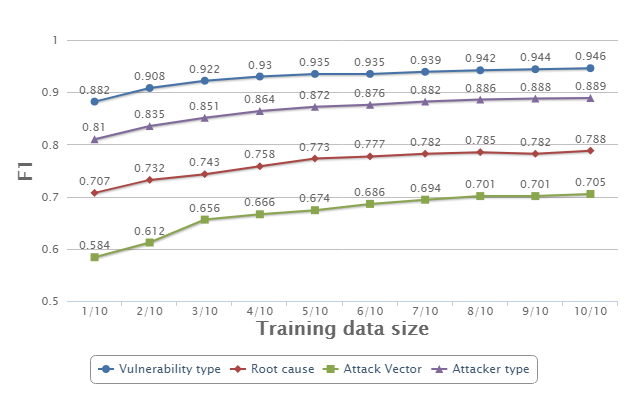}
	\vspace{-1mm}
	\caption{Impact of training data size on F1-score}
	\vspace{-2mm}
	\label{fig:split_data}
\end{figure}

\subsection{Size of Training Data (RQ3)}

\noindent
\textbf{Motivation:}
The prediction our approach makes is based on the correlations between a to-be-predicted aspect and the other aspects learned from the descriptions of historical CVEs.
To understand the practicality of our approach, we investigate the minimal data required for training a reliable prediction model, and how the prediction performance changes as the training data increases.

\vspace{0.5mm}
\noindent
\textbf{Approach:}
Again, we use the most effective classifier design identified in RQ1.
For each to-be-predicted aspect, we randomly split its training dataset into 10 equal-sized subsets.
We train 10 classifiers with $n/10$ ($1 \leq n \leq 10$) training data respectively.
The training data starts with one randomly selected subset.
For each increment, we randomly select one of the remaining subsets and add it to the existing training data.
The 10 classifiers are tested on the same set of testing data.
We compute the evaluation metrics for the $n/10$ training data size.
We perform 10-fold cross validation and obtain the average metrics for each increment.

\vspace{0.5mm}
\noindent
\textbf{Results:}
Figure~\ref{fig:split_data} shows the F1 results at different training data sizes.
As the size of training data increases, the prediction performance improves.
For Vulnerability type aspect, the performance improvement is statistically significant from 1/10 to 4/10 training data.
However, from 5/10 training data onwards, increasing the training data further results in only statistically non-significantly performance improvement.
For the attacker type aspect and impact aspect, the performance improvement is statistically significant from 1/10 to 5/10 training data.
However, from 6/10 training data onwards, increasing the training data further results in only statistically non-significantly performance improvement.
For the root cause aspect, the performance improvement of training data from 1/10 to 8/10 was statistically significant.
As the size of root-cause dataset is about half of the size of the other three aspects (see Table~\ref{tab:dataset}), the amount of CVEs required for training a reliable root-cause classifier is similar to the other three aspects. 

\noindent\fbox{\begin{minipage}{8.4cm} \emph{Our approach is practical as it requires only about 17,500-21,000 historical CVEs (less than 20\% of all available CVEs) to train a reliable classifier for predicting various missing CVE aspects.}
\end{minipage}}\\

\subsection{Results: Ablation Study (RQ4)}
\label{sec:aspectablation}
\begin{table}
	\setlength{\abovecaptionskip}{-0cm}%
	\setlength{\belowcaptionskip}{0cm}%
	\scriptsize
	\centering
	\caption{\label{tab:reduce_vt}Ablation results for predicting vulnerability type}
\setlength{\tabcolsep}{0.9mm}{
	\begin{tabular}{ccccccc}
		\toprule
		&Ablated aspect&Root cause&Affected product& Impact&Attacker type &Attack vector\\
		\midrule
		
		&Precision & 0.943    & 0.925&  0.821 &0.939 & 0.888\\
		&Recall & 0.943 & 0.927& 0.822  &0.941& 0.896  \\
		
		&F1  & 0.943& 0.925& 0.821 &0.939 &0.890\\

		\bottomrule
	\end{tabular}
}
\end{table}

\begin{table}
\vspace{-1mm}
	\setlength{\abovecaptionskip}{0.2cm}%
	\setlength{\belowcaptionskip}{0.1cm}%
	\scriptsize
	\centering
	\caption{\label{tab:reduce_rc}Ablation results for predicting root cause}
	\vspace{-2mm}
\setlength{\tabcolsep}{0.9mm}{
	\begin{tabular}{ccccccc}
		\toprule
		&Ablated aspects& Vul-type&Affected product & Impact&Attacker type  &Attack vector\\
		\midrule
		&Precision  & 0.740 &0.734&  0.739 &0.781&0.780\\

		&Recall  & 0.751 &0.741 & 0.755&0.793&0.795  \\

		&F1  & 0.745& 0.730& 0.736  &0.785&0.784\\
		\bottomrule
	\end{tabular}
}\vspace{-2mm}
\end{table}

\begin{table}
	\scriptsize
	\centering
	\setlength{\abovecaptionskip}{0.2cm}%
	\setlength{\belowcaptionskip}{0.1cm}%
	\caption{\label{tab:reduce_at}Ablation results for predicting attacker type}
	\vspace{-2mm}
	\setlength{\tabcolsep}{0.9mm}{
		\begin{tabular}{ccccccc}
			\toprule
			&Ablated aspect & Vul-type & Root cause&Affected product& Impact&Attack vector \\
			\midrule
			&Precision    &  0.852&0.873  &0.850& 0.883& 0.864  \\
			
			&Recall     & 0.876 &0.892 &0.874& 0.895 &0.871  \\
			
			&F1  & 0.861&0.878&0.847  & 0.881 & 0.863 \\

			\bottomrule
		\end{tabular}
	}
	\vspace{-2mm}
\end{table}

\begin{table}
	\vspace{-1mm}
	\scriptsize
	\setlength{\abovecaptionskip}{0.2cm}%
	\setlength{\belowcaptionskip}{0.1cm}%
	\centering
	\caption{\label{tab:reduce_av}Ablation results for predicting attack vector}
	\vspace{-2mm}
\setlength{\tabcolsep}{0.9mm}{
	\begin{tabular}{ccccccc}
		\toprule
		&Ablated aspect & Vul-type& Root cause &Affected product& Impact&Attacker type \\
		\midrule
		&Precision    &  0.659&0.696 & 0.568 & 0.680 &0.670\\
		
		&Recall     & 0.693 &0.701&0.601 & 0.700 &0.674 \\
		
		&F1  & 0.665  &0.695& 0.572 & 0.683&0.669\\

		\bottomrule
	\end{tabular}
}
	\vspace{-2mm}
\end{table}

\textbf{Motivation:}
As shown in Table~\ref{tab:dataset}, it is unrealistic to assume that all other five CVE aspects are available for predicting a particular aspect.
In this RQ, we want to investigate the impact of certain CVE aspects unavailable as input on the accuracy of predicting a particular aspect.
This study has two important roles.
First, it identifies stronger correlations (if any) among some CVE aspects than others.
Second, it identifies the minimum subset of known aspects required for making reliable prediction of a particular aspect. 
This also help us to understand the practicality of our approach.

\textbf{Approach:}
For a particular aspect, we conduct five experiments.
In each experiment, we ablate one of the other five aspects in the aspect-specific dataset, which produces an ablation dataset without the ablated aspect.
For example, for vulnerability type, we obtain five datasets without root cause, affected product, impact, attacker type or attack vector for the five ablation experiments, respectively.
Although the experiments in RQ1/RQ2/RQ3 do not assume the availability of all five aspects, the ablation of one aspect in this RQ means that we completely ignore this ablated aspect as known aspect for model input, even the CVEs describe this ablated aspect.
We use the most effective classifier design identified in RQ1, and train and test the classifier on each ablation dataset.
We perform 10-fold cross-validation in all experiments.

\textbf{Results:}
Table~\ref{tab:reduce_vt}-Table~\ref{tab:reduce_av} show our experimental results. 
We compare the performance metrics with those by 1-layer CNN in Table~\ref{tab:detail}.
For predicting vulnerability type, ablating impact results in the most significant drop (12.6\%) in F1, followed by ablating attack vector (5.6\% drop in F1).
In contrast, ablating attacker type and affected product have a much less significant impact, with 0.7\% and 2.1\% drop in F1 respectively. 
Ablating root cause almost has no impact on predicting vulnerability type.
As discussed in Section~\ref{sec:extractionquality}, over 94\% of CVEs describe impact aspect.
Therefore, our approach will not actually suffer from the performance degradation due to the unavailability of impact as input in practice.
Although unavailable attack vector as input aspect affects the prediction of vulnerability type, our approach can still achieve high accuracy (0.89 in F1).

For predicting attack vector, ablating affected product has the most significant impact, resulting in 13.4\% drop in F1.
However, as almost all CVEs describe affected product, our approach will not actually suffer from the performance degradation due to the unavailability of affected product as input.
Ablating the other four aspects results in much smaller drop (about 1.1\%-4.1\%) in F1.
Among these four aspects, vulnerability type and attacker type have stronger correlations with attack vector than impact and root cause.

There are no such a prominent aspect for predicting root cause and attacker type as impact for vulnerability type and affected product for attack vector.
For predicting root cause, ablating vulnerability type, affected product or impact has relatively larger impact (about 5\% drop in F1), compared with about 0.4\% drop in F1 by ablating attacker type and attack vector.
For predicting attacker type, ablating affected product results in relative larger drop in F1 (4.2\%) than ablating the other four aspects.
Again, we do not have the real issue of unavailable affected product and impact as input in practice.

\noindent\fbox{\begin{minipage}{8.4cm} \emph{Our ablation study identifies two strong correlations: vulnerability type and impact, affected product and attack vector. Although the unavailability of certain aspect(s) in the CVE description degrades the prediction performance of an unknown aspect, it does not significantly affect the practicality of our approach.} \end{minipage}}\\


\section{Threats to Validity}


Our approach relies on the patterns of CVE descriptions, in terms of what aspects people describe in the vulnerability reports and how they describe these aspects.
This assumption holds in general because there is a common knowledge about important aspects of vulnerabilities and much effort has been made to standardize the description of these aspects~\cite{poj,cvesubmissionform}.
Therefore, patterns exist in historical CVEs, which can be learned by an appropriate machine learning methods.
Our approach currently uses a rule-based method to extract CVE aspects from CVE descriptions.
We obtain aspect extraction rules by observing about 24000 CVEs from 1999 to 2019, and confirm the extraction accuracy of these rules.
However, the development of aspect extraction rules may suffer from human biases and errors, and the developed rules may not cover the emerging CVE description patterns.
We are now investigating machine-learning based semantic role labeling method to address this limitation.
Finally, although our study demonstrates the practicality of predicting missing aspects of vulnerability reports, how to integrate our approach in CVE submission process to improve the information completeness of CVEs requires future study.

\section{Related Work}

Various databases have been created to document and analyze publicly know software vulnerabilities.
For example, Common Vulnerabilities and Exposures (CVE)~\cite{cve} and SecurityFocus~\cite{sf} are two well known vulnerability databases. 
Common Weakness Enumeration (CWE) abstract common software weaknesses of individual vulnerabilities, which are often referred to as vulnerability type of CVEs.
New software vulnerabilities have been regularly discovered and added to the vulnerability database.
For example, in about six months from November 2019 to May 2020, 5,685 new CVEs have been added to the CVE database.

The fast growing vulnerabilities demand automatic methods to assist the analysis of the properties of newly discovered vulnerabilities.
As newly discovered vulnerabilities often exhibit similar properties as historical vulnerabilities, machine learning approaches have been proposed for predicting vulnerability properties.
For example, Bozorgi et al.~\cite{c18} train a classifier based on various features in vulnerability reports, such as description and time stamp, to predict the exploitability of a vulnerability.
Han et al.~\cite{han2017learning} propose a CNN-based classifier to predict the severity of CVEs based on only the CVE description.
Gong et al.~\cite{gong2019joint} develop a multi-task learning method to predict seven vulnerability properties according to the Common Vulnerability Scoring System~\cite{cvss}.
Xiao et al.~\cite{xiao2019embedding} construct a knowledge graph of CVEs and CWEs, and propose a graph embedding method to infer the relationships of software vulnerabilities and weaknesses.
Different from these works, our work studies the information completeness of vulnerability reports and develops a neural network classifier for predicting the missing information of key aspects in the vulnerability reports. 

Neural network based techniques have been widely adopted for text classification in natural language community~\cite{kim2014convolutional, feldman2013techniques, Matthew2018Deep}.
They have also been applied to software text other than vulnerability descriptions, as well as source code.
For example, Xu et al.~\cite{c28} develop a CNN-based siamese network to predict duplicate questions on Stack Overflow.
Chen et al.~\cite{c20} develop a similar Siamese network architecture but their goal is to support cross-lingual question retrieval.
Li et al.~\cite{c21} develop QDLinker based on word embeddings and CNN for answering programming questions with software documentation.
Mou et al.~\cite{mou2016convolutional} propose a tree-based CNN to embed source code for classifying programs and detecting code patterns. 

Much research have been done on predicting vulnerable or error-prone components~\cite{c13,c14}, or assess in which ways a system is more likely to be attacked\cite{c17}.
They use various features including software metrics, code churn, developer activity metrics, code structure~\cite{c15,c16}.
Different from these works, our work analyzes vulnerability text and learn correlations between different aspects in the vulnerability descriptions.

\section{Conclusion and Future Work}

This paper studies the information missing issue in the vulnerability reports.
We examine six key aspects of CVE descriptions and find different severities of information missing for root cause, vulnerability type, attacker type and attack vector.
We propose a machine learning approach for completing the missing information of these four aspects in the CVE descriptions.
Our approach uses neural network model to extract important features from aspect descriptions and capture intrinsic correlations among different aspects.
Our large-scale experiments identify the most effective model design for the prediction task.
Our model can be trained effectively using historical CVEs, and the trained model can accurately predict missing information of CVE aspects, which could alleviate the information missing issue in the vulnerability reports.
Our experiments on the minimum effective amount of training data and the prominent correlations among different aspects confirm the practicality of our approach for completing missing information of key aspects in vulnerability reports.
\newpage\newpage
\bibliographystyle{ACM-Reference-Format}
\bibliography{references}

\end{document}